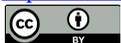

# Do the Amati and Yonetoku Relations Evolve with Redshift for *Swift* GRBs?

**Ali M. Hasan, Walid J. Azzam**

Department of Physics, College of Science, University of Bahrain, Sakhir, Bahrain
Email: wjazzam@uob.edu.bh, wjazzam@gmail.com

## Abstract

Gamma-ray bursts (GRBs) are extremely powerful stellar explosions that have been observed to huge distances with redshifts exceeding 9. Although GRBs are not standard candles, one may "standardize" them by calibrating certain correlations that link an intrinsic parameter to an observed one. Two such correlations that have been discovered are the Amati relation and the Yonetoku relation. The former is a correlation between a burst's equivalent isotropic energy, $E_{iso}$, and its intrinsic peak spectral energy $E_{p,i}$, while the latter is a correlation between a burst's peak isotropic luminosity, $L_{iso}$, and $E_{p,i}$. In this paper, we compiled a large sample of 241 *Swift* long GRBs for the purpose of examining whether the Amati and Yonetoku relations are immune to redshift evolution. Our methodology encompasses two approaches: the first involves binning the data by redshift and fitting the two relations for each bin, then checking whether the fitting parameters evolve with redshift; the second approach involves using a redshift cutoff to divide the data into a low-redshift group and a high-redshift group, then checking whether the fitting parameters for the two relations are consistent with one another. Our results indicate that the Amati and Yonetoku relations are robust in the sense that they do not show any systematic or significant redshift evolution. Moreover, our results indicate that the high redshift bins show better fits compared to the low redshift bins, which indicates that the Amati and Yonetoku relations are more reliable as "standard candles" for high redshift and hence are promising cosmological probes.

## Keywords

Gamma-Ray Bursts, Spectral Correlations, Redshift Evolution, Cosmological Probes

## 1. Introduction

Gamma-ray Bursts (GRBs) are extremely powerful explosions that have been ob-served to very high redshift, which makes them particularly suitable for early universe investigations [1]-[3]. Although several studies have highlighted their potential as "standard candles" in cosmology [4]-[7], there are still a few enigmatic issues regarding their origins and properties. To resolve some of these issues, several studies examined the intrinsic characteristics of GRBs and found several important correlations among GRB parameters. One such correlation is the Amati relation, which links the equivalent isotropic energy $E_{iso}$ of a burst to its intrinsic peak spectral energy $E_{p,i}$ through a power-law correlation, $E_{iso} \propto E_{p,i}^m$ [8]-[11]. As several studies stress, the Amati relation is important because it is a potential tool for constraining cosmological parameters [4] [5] [7] [12]-[16]. In addition, some studies argue that it can be used to classify GRBs. Traditionally, GRBs are classified into long GRBs (LGRBs) with $T_{90} > 2$ s and short GRBs (SGRBs) with $T_{90} < 2$ s, where $T_{90}$ is the time taken to observe 90% of the GRB's fluence [17]. However, this classification is not based on the intrinsic properties, making it prone to cosmological effects [18]-[22]. Recently, many studies have explored other methods of classification like employing the Amati relation as a discriminator among different GRB classes. More specifically, some recent studies have found evidence that SGRBs and LGRBs have different Amati slopes, which means that the Amati relation may be used to distinguish the two classes [23]-[25].

Another important correlation is the Yonetoku relation. The Yonetoku relation connects the peak isotropic luminosity $L_{iso}$ to the intrinsic peak spectral energy $E_{p,i}$ through a power law as well, $L_{iso} \propto E_{p,i}^m$ [26]-[28]. Although some studies have utilized the Yonetoku relation as a distance indicator [7] [29], it is important to caution that to use the Yonetoku relation, or for that matter the Amati relation, as a cosmological probe, it must be calibrated properly. One problem that arises regarding the calibration of the Amati and Yonetoku relations is whether these correlations evolve with redshift. Both the Amati relation and the Yonetoku relation have been the subject of several studies that investigated their potential redshift evolution. Early studies found varying results regarding the evolution of these two relations [13] [28] [30]-[36]. For instance, a study in 2013 found evidence that the fitting slopes for both the Amati and Yonetoku relations do not evolve with redshift [37]. However, these studies suffered from the paucity of data points, as not many GRBs observed, at the time, had available redshift and spectral parameters.

As more input data have become available, there has been a revived interest in exploring whether intrinsic GRB correlations, like the Amati and Yonetoku relations, evolve with redshift. For instance, the recent study by Jia *et al.* [16], uses 221 GRBs to study the redshift evolution of the Amati relation. They split the dataset into 5 redshift bins and then fit each bin with the Amati relation. They find that the results of all the bins are consistent with one another within $2\sigma$, suggesting no redshift evolution. On the other hand, two other studies, Singh *et al.* [38] and Singh *et al.* [39], find evidence of redshift evolution in the Amati relation. These two studies split their GRB datasets into low and high redshift subsets using a redshift cutoff and fit each data subset using the Amati relation. In [38], a redshift cutoff of 1.5 is used for a GRB dataset composed of 162 bursts, and the authors find that their two subsets show an inconsistency in the Amati relation at the $2\sigma$ level, hinting at a possible redshift evolution. In the second study [39], the authors use the same dataset employed by [37] and carry out an analysis like that conducted by [38], and they find that again there is evidence, at the $2\sigma$ level, that the Amati relation evolves with redshift. On the other hand, less attention has been given in the past few years regarding the potential redshift evolution of the Yonetoku relation.

Given the importance of these two relations, we aim in this study to carry out a systematic investigation of their potential redshift evolution using a large dataset of 241 *Swift* LGRBs. In Section 2, we describe our data sample along with our methodology, and we present and discuss our results in Section 3. We provide a brief conclusion in Section 4.

## 2. Data Sample and Methodology

We compiled our GRB data sample from the *Swift* catalog [40]. As of July 2025, *Swift* has detected 1717 GRBs, of which 453 have reported redshifts. To fit the purpose of our study, we selected GRBs that satisfied the following criteria:

- The GRB must have a $T_{90} > 2$ s, in other words, it must be a long burst according to the traditional classification of GRBs [15] (27 GRBs removed).
- The GRB must have a single value detected redshift (12 GRBs removed).
- The GRB must have at least an estimated fluence $S_{obs}$, a 1-sec peak photon flux $F_{\gamma}$, along with time-averaged and 1-sec peak spectral parameters. Note that in the catalog, only the power-law and the cutoff power-law fittings are reported and for our purpose, we use the latter as it provides an estimate of the peak energy $E_p$ (153 GRBs removed).
- We excluded GRBs with large errors: exceeding 500% in their fluence, peak photon flux, or peak energy (20 GRBs removed).

This left us with 241 LGRBs suitable for our investigation. In addition, to account for the completeness of our dataset, we also compiled a subsample in which the bursts had a 1-sec peak photon flux $F_{\gamma} \geq 2.6$ photons/cm²/s. This new subset, which we will call the filtered dataset, has 87 LGRBs. For comparison reasons and to check the completeness of our original sample, we will report and compare the results of both datasets.

The redshift distribution of both datasets is shown in **Figure 1**. The redshift distribution of the unfiltered dataset is comparable to what we found in a previous study that we conducted [41], implying an excess of GRBs at low redshift compared to the star-formation rate [41]. Filtering the dataset by imposing the $F_{\gamma} \geq 2.6$ photons/cm²/s condition removes a substantial number of high redshifts GRBs, leading to more excess GRBs at low redshift (when the distribution is normalized).

We wrote the Amati and Yonetoku relations in their logarithmic form and then carried out a linear fit:

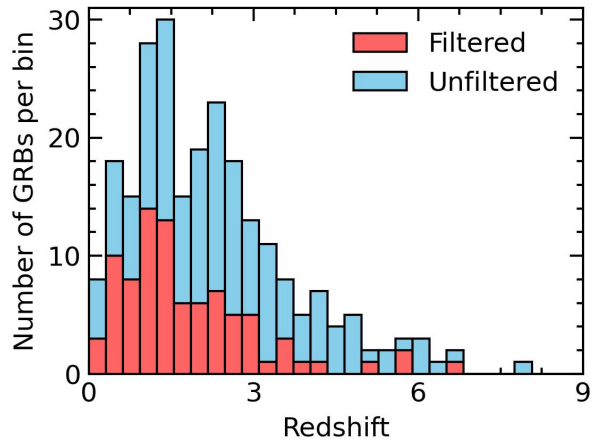

**Figure 1.** The redshift distribution of the datasets, with a bin size of 0.3.

$$\log\left(\frac{E_{iso}}{10^{52}\,\mathrm{erg}}\right) = A\log\left(\frac{E_{p,i}}{1\,\mathrm{keV}}\right) + B, \tag{1}$$

$$\log\left(\frac{L_{iso}}{10^{52}\,\mathrm{erg/s}}\right) = A'\log\left(\frac{E_{p,i}}{1\,\mathrm{keV}}\right) + B', \tag{2}$$

where $E_{p,i} = E_p^{\mathrm{obs}}(1+z)$, is the intrinsic peak energy. $E_{iso}$ and $L_{iso}$ were calculated using the $k$-correction introduced by Bloom *et al.* (2001) [42] and implemented as detailed by Zitouni *et al.* (2014) [43]. Note that the *Swift* energy band is [15 - 150] keV and we considered the rest-frame energy band to be [1 - $10^4$] keV for the k-correction method. For the cosmological parameters, we adopted the ΛCMD model with a Hubble constant $H_0 = 70.8$ km/s/Mpc, $\Omega_M = 0.27$, and $\Omega_\Lambda = 1 - \Omega_M = 0.73$.

We employed two approaches to investigate the redshift evolution of the Amati and Yonetoku relations. First, a binning approach was used where the datasets were sorted according to redshift and then split into 3, 4 and 5 bins with an equal number of GRBs. Each bin was fitted individually with both relations. The second approach involved splitting the dataset into two groups, a low-redshift group and a high-redshift group, using a redshift cutoff. Each group was fitted with both relations. This method was used by Singh *et al.* [38] and Singh *et al.* [39], and following a similar procedure as the latter, we tried redshift cutoffs of 1.5 and 2.0 to split our datasets.

We used a least $\chi^2$ method to find the best fit parameters, where $\chi^2$ is calculated as follows:

$$\chi^2 = \sum_i \frac{\left(y_i^{\mathrm{obs}} - y_i^{\mathrm{calc}}\right)^2}{\sigma_{\mathrm{tot}}^2}, \tag{3}$$

with $\sigma_{\mathrm{tot}}^2 = \sigma_{\mathrm{obs}}^2 + \sigma_{\mathrm{int}}^2$. Here, $\sigma_{\mathrm{int}}$ is the intrinsic scattering parameter which encapsulates other errors not accounted for by the observational error. The optimization is run for parameters A, B and $\sigma_{\mathrm{int}}$. In general, a smaller $\sigma_{\mathrm{int}}$ means that our model and the data fit well. This method is equivalent to maximizing the like-lihood function $\mathcal{L}$, which is related to $\chi^2$ as follows:

$$\chi^2 = -2\ln(\mathcal{L}).\qquad(4)$$

For all our fits, we report the fitting parameters, the intrinsic scattering parameter, and the Akaike Information Criterion (AIC). The AIC is calculated from the likelihood function $\mathcal{L}$ or the $\chi^2$ value in the following way [44] [45]:

$$\text{AIC} = 2k + \chi^2 = 2k - 2\ln(\mathcal{L}),\qquad(5)$$

with $k$ being the number of fitting parameters. Note that AIC is important when comparing datasets with different numbers of parameters, such as the binned datasets.

## 3. Results and Discussion

Table 1 provides a summary of calculated parameters and compares the unfiltered and filtered datasets. We note that the filtered dataset removes a good portion of low luminosity GRBs but also removes some high redshift GRBs, which are primarily high luminosity. This leads to a lower mean $L_{iso}$.

Table 1. A general description for the redshift, $E_{p,i}$, $E_{iso}$ and $L_{iso}$ for the unfiltered and filtered datasets.

| Dataset | | Redshift | $E_{p,i}$ (keV) | $E_{iso}$ (ergs) | $L_{iso}$ (ergs/s) |
|---|---|---|---|---|---|
| Unfiltered | Range | 0.0389 - 8 | 16.56 - 4324.77 | $3.85 \times 10^{48}$ - $6.04 \times 10^{54}$ | $1.36 \times 10^{48}$ - $9.65 \times 10^{55}$ |
| | Mean | 2.23 | 439.95 | $2.91 \times 10^{53}$ | $5.01 \times 10^{53}$ |
| Filtered | Range | 0.13 - 6.782 | 37.53 - 3404.88 | $3.82 \times 10^{50}$ - $6.04 \times 10^{54}$ | $1.08 \times 10^{50}$ - $3.31 \times 10^{54}$ |
| | Mean | 1.81 | 496.95 | $4.01 \times 10^{53}$ | $1.65 \times 10^{53}$ |

Table 2. The Amati and Yonetoku fitting results for the unbinned datasets.

| Relation | A | B | $\sigma_{int}$ | AIC |
|---|---|---|---|---|
| | Unfiltered dataset | | | |
| Amati | $1.28 \pm 0.15$ | $-2.73 \pm 0.36$ | $0.82 \pm 0.02$ | 231.33 |
| Yonetoku | $2.24 \pm 0.14$ | $-5.69 \pm 0.36$ | $0.52 \pm 0.01$ | 172.26 |
| | Filtered dataset ($F_\gamma \geq 2.6$ photons/cm²/s) | | | |
| Amati | $1.56 \pm 0.22$ | $-3.23 \pm 0.57$ | $0.70 \pm 0.24$ | 86.12 |
| Yonetoku | $2.20 \pm 0.16$ | $-5.32 \pm 0.43$ | $0.42 \pm 0.03$ | 74.41 |

Table 2 shows our Amati and Yonetoku fitting results for the unbinned datasets. We note that both datasets, the filtered and the unfiltered, are consistent with the Amati and Yonetoku relations, though the Yonetoku relation shows a better fit. Comparing the Amati parameters of the filtered and unfiltered datasets, we see that there is a difference between them, with the unfiltered results agreeing more with the results reported in [8], while the result of the filtered dataset is closer to that found by [39]. This then suggests that the Amati relation is prone to issues pertaining to the completeness of the dataset. As for the Yonetoku relation, we find both datasets are consistent with each other, and they agree with the values reported in the literature [26].

**Table 3.** The Amati and Yonetoku fitting results for the binned unfiltered dataset.

| Number of bins | Bin (redshift) | A | B | $\sigma_{int}$ | AIC | Total AIC |
|---|---|---|---|---|---|---|
| | | Amati relation | | | | |
| | Bin 1 (0.85 ± 0.36) | 0.84 ± 0.35 | −2.18 ± 0.81 | 0.98 ± 0.20 | 70.92 | |
| 3 bins | Bin 2 (1.97 ± 0.36) | 1.08 ± 0.20 | −2.04 ± 0.50 | 0.70 ± 0.01 | 75.90 | 224.85 |
| | Bin 3 (3.89 ± 1.18) | 0.87 ± 0.22 | −1.40 ± 0.57 | 0.60 ± 0.13 | 74.03 | |
| | Bin 1 (0.71 ± 0.30) | 0.40 ± 0.43 | −1.32 ± 0.99 | 1.06 ± 0.23 | 49.55 | |
| 4 bins | Bin 2 (1.53 ± 0.24) | 1.14 ± 0.30 | −2.30 ± 0.75 | 0.76 ± 0.16 | 50.79 | 214.61 |
| | Bin 3 (2.44 ± 0.26) | 1.05 ± 0.20 | −1.98 ± 0.51 | 0.61 ± 0.01 | 56.63 | |
| | Bin 4 (4.27 ± 1.12) | 0.89 ± 0.26 | −1.40 ± 0.69 | 0.62 ± 0.14 | 57.64 | |
| | Bin 1 (0.62 ± 0.26) | 0.39 ± 0.52 | −1.46 ± 1.17 | 1.12 ± 0.28 | 34.65 | |
| | Bin 2 (1.30 ± 0.15) | 0.70 ± 0.30 | −1.32 ± 0.74 | 0.75 ± 0.57 | 43.99 | |
| 5 bins | Bin 3 (1.98 ± 0.24) | 2.13 ± 0.28 | −4.61 ± 0.72 | 0.55 ± 0.02 | 46.73 | 221.24 |
| | Bin 4 (2.74 ± 0.26) | 0.75 ± 0.19 | −1.22 ± 0.49 | 0.57 ± 0.17 | 49.17 | |
| | Bin 5 (4.56 ± 1.07) | 0.89 ± 0.32 | −1.38 ± 0.86 | 0.63 ± 0.15 | 46.70 | |
| | | Yonetoku relation | | | | |
| | Bin 1 (0.85 ± 0.36) | 1.54 ± 0.34 | −4.35 ± 0.82 | 0.67 ± 0.06 | 65.76 | |
| 3 bins | Bin 2 (1.97 ± 0.36) | 2.06 ± 0.18 | −5.07 ± 0.49 | 0.34 ± 0.01 | 52.14 | 163.96 |
| | Bin 3 (3.89 ± 1.18) | 1.88 ± 0.19 | −4.41 ± 0.53 | 0.23 ± 0.01 | 46.06 | |
| | Bin 1 (0.71 ± 0.30) | 1.15 ± 0.40 | −3.55 ± 0.96 | 0.71 ± 0.16 | 49.17 | |
| 4 bins | Bin 2 (1.53 ± 0.24) | 1.93 ± 0.23 | −4.83 ± 0.59 | 0.33 ± 0.01 | 46.32 | 167.60 |
| | Bin 3 (2.44 ± 0.26) | 1.99 ± 0.21 | −4.72 ± 0.58 | 0.30 ± 0.01 | 38.49 | |
| | Bin 4 (4.27 ± 1.12) | 2.25 ± 0.16 | −5.53 ± 0.49 | 0.00 ± 0.00 | 33.62 | |
| | Bin 1 (0.62 ± 0.26) | 1.10 ± 0.46 | −3.52 ± 1.09 | 0.77 ± 0.30 | 37.92 | |
| | Bin 2 (1.30 ± 0.15) | 1.86 ± 0.25 | −4.77 ± 0.66 | 0.37 ± 0.01 | 36.76 | |
| 5 bins | Bin 3 (1.98 ± 0.24) | 2.24 ± 0.25 | −5.49 ± 0.67 | 0.36 ± 0.03 | 34.52 | 165.35 |
| | Bin 4 (2.74 ± 0.26) | 1.72 ± 0.25 | −3.98 ± 0.69 | 0.28 ± 0.02 | 28.77 | |
| | Bin 5 (4.56 ± 1.07) | 2.27 ± 0.18 | −5.59 ± 0.54 | 0.00 ± 0.00 | 27.39 | |

Our results for the unfiltered and filtered binned datasets are presented in Table 3 and Table 4, respectively. A visual comparison between the parameters is shown in Figure 2 and Figure 3, for the unfiltered and filtered dataset, respectively. For the Amati relation, we see that the unfiltered dataset shows no redshift evolution, with all the bins being within at most $2\sigma$ from each other and from the full dataset fitting parameters. The filtered dataset shows a trend when moving from low to high redshift, especially for 5 bins, where the high redshift bins show strong agreement with the fitting of the whole dataset. But one cannot claim there is a redshift evolution here due to the large errors where all the bins are within a $1\sigma$ range. The different behavior between the two datasets further suggests that the Amati relation is affected by the completeness of the dataset. Overall, the Amati relation does not show clear signs of redshift evolution. The Yonetoku relation, in a similar manner, does not show any redshift evolution behavior with all bins being within $2\sigma$. However, the high redshift bins appear to be more consistent and show better fitting than low redshift ones, which is witnessed by comparing them to the whole dataset parameters and is also reflected in the $\sigma_{int}$ trends. In both datasets, the $\sigma_{int}$ value in the high redshift bins is smaller than the low redshift bins, suggesting that it fits the model better. In other words, the Yonetoku relation is more reliable for high redshift GRBs than for low redshift ones.

**Table 4.** The Amati and Yonetoku fitting results for the binned filtered dataset.

| Number of bins | Bin (redshift) | A | B | $\sigma_{int}$ | AIC | Total AIC |
|---|---|---|---|---|---|---|
| | | | Amati relation | | | |
| | Bin 1 (0.85 ± 0.36) | 0.56 ± 0.53 | −1.25 ± 1.20 | 0.67 ± 2.18 | 31.59 | |
| 3 bins | Bin 2 (1.97 ± 0.36) | 1.35 ± 0.57 | −2.53 ± 1.47 | 0.89 ± 0.51 | 23.77 | 84.37 |
| | Bin 3 (3.89 ± 1.18) | 1.25 ± 0.31 | −2.20 ± 0.86 | 0.49 ± 0.16 | 29.00 | |
| | Bin 1 (0.71 ± 0.30) | 0.51 ± 0.67 | −1.24 ± 1.54 | 0.70 ± 0.26 | 18.56 | |
| 4 bins | Bin 2 (1.53 ± 0.24) | 0.46 ± 0.46 | −0.66 ± 1.14 | 0.70 ± 0.37 | 23.70 | 81.58 |
| | Bin 3 (2.44 ± 0.26) | 1.89 ± 0.39 | −3.70 ± 1.06 | 0.52 ± 0.12 | 20.80 | |
| | Bin 4 (4.27 ± 1.12) | 1.00 ± 0.47 | −1.53 ± 1.32 | 0.62 ± 0.31 | 18.51 | |
| | Bin 1 (0.62 ± 0.26) | 0.09 ± 0.80 | −0.40 ± 1.81 | 0.77 ± 0.31 | 13.25 | |
| | Bin 2 (1.30 ± 0.15) | 0.61 ± 0.64 | −1.07 ± 1.54 | 0.61 ± 0.24 | 21.46 | |
| 5 bins | Bin 3 (1.98 ± 0.24) | 0.81 ± 0.58 | −1.18 ± 1.49 | 0.78 ± 0.35 | 17.89 | 85.41 |
| | Bin 4 (2.74 ± 0.26) | 1.32 ± 0.39 | −2.25 ± 1.08 | 0.52 ± 0.15 | 18.82 | |
| | Bin 5 (4.56 ± 1.07) | 1.48 ± 0.67 | −2.84 ± 1.88 | 0.69 ± 0.34 | 14.00 | |
| | | | Yonetoku relation | | | |
| | Bin 1 (0.85 ± 0.36) | 0.95 ± 0.47 | −2.73 ± 1.10 | 0.48 ± 0.10 | 25.75 | |
| 3 bins | Bin 2 (1.97 ± 0.36) | 2.09 ± 0.24 | −4.96 ± 0.65 | 0.28 ± 0.03 | 26.56 | 72.85 |
| | Bin 3 (3.89 ± 1.18) | 1.65 ± 0.18 | −3.50 ± 0.52 | 0.16 ± 0.01 | 20.54 | |
| | Bin 1 (0.71 ± 0.30) | 1.16 ± 0.69 | −3.33 ± 1.60 | 0.54 ± 0.19 | 18.61 | |
| 4 bins | Bin 2 (1.53 ± 0.24) | 1.65 ± 0.22 | −3.96 ± 0.57 | 0.23 ± 0.02 | 18.77 | 74.63 |
| | Bin 3 (2.44 ± 0.26) | 2.10 ± 0.26 | −4.86 ± 0.73 | 0.27 ± 0.04 | 21.42 | |
| | Bin 4 (4.27 ± 1.12) | 1.58 ± 0.25 | −3.28 ± 0.71 | 0.15 ± 0.01 | 15.82 | |
| | Bin 1 (0.62 ± 0.26) | 0.76 ± 0.77 | −2.56 ± 1.79 | 0.54 ± 0.20 | 13.16 | |
| | Bin 2 (1.30 ± 0.15) | 0.86 ± 0.32 | −2.13 ± 0.79 | 0.15 ± 0.03 | 16.77 | |
| 5 bins | Bin 3 (1.98 ± 0.24) | 1.84 ± 0.24 | −4.32 ± 0.63 | 0.18 ± 0.02 | 15.62 | 76.98 |
| | Bin 4 (2.74 ± 0.26) | 1.82 ± 0.22 | −3.92 ± 0.63 | 0.20 ± 0.06 | 19.27 | |
| | Bin 5 (4.56 ± 1.07) | 1.76 ± 0.27 | −3.87 ± 0.84 | 0.00 ± 0.00 | 12.16 | |

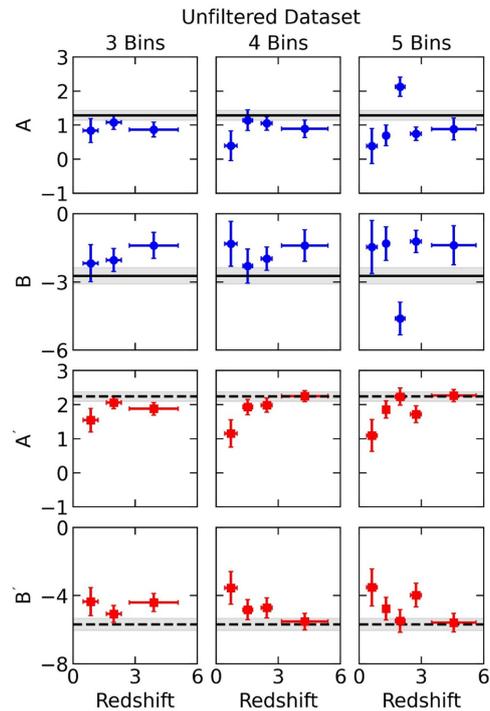

**Figure 2.** The fitting results for the binned unfiltered dataset shown for 3, 4, and 5 bins. The values for the fitting parameters, A and B, for the Amati and Yonetoku relations are shown in blue and red, respectively. For comparison, the horizontal lines, solid and dashed, show the values of A and B when the entire unfiltered dataset is used (*i.e.*, without binning).

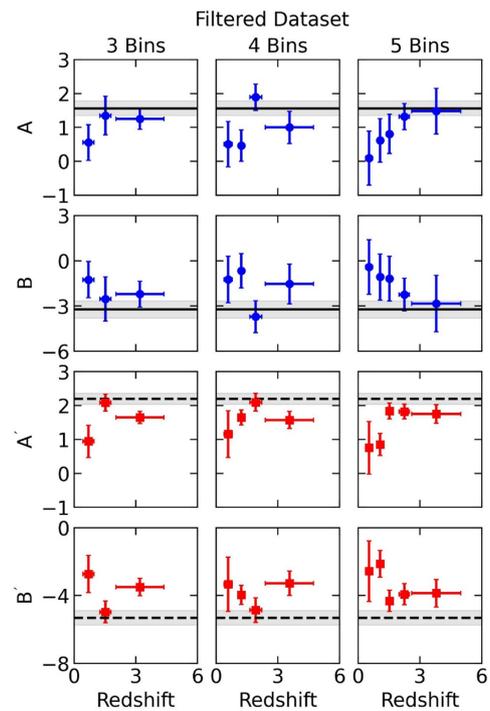

**Figure 3.** The fitting results for the binned filtered dataset shown for 3, 4, and 5 bins. The values for the fitting parameters, A and B, for the Amati and Yonetoku relations are shown in blue and red, respectively. For comparison, the horizontal lines, solid and dashed, show the values of A and B when the entire filtered dataset is used (*i.e.*, without binning).

Table 5 presents our results for the cutoff splitting of both datasets. We can see that the results from both datasets agree with the parameters found from fitting the whole datasets (Table 2), within at most $2\sigma$. Again, we find that the high redshift groups, $z \geq z_{\text{cutoff}}$, show better fits and are closer to the whole dataset fitting than the low redshift groups, $z < z_{\text{cutoff}}$ groups.

**Table 5.** The Amati and Yonetoku relations fitting results for the datasets that were split into groups using a redshift cutoff.

| Relation | Group | A | B | $\sigma_{\text{int}}$ | AIC | Total AIC |
|---|---|---|---|---|---|---|
| | | | Unfiltered dataset | | | |
| | | | Cutoff $z_{\text{cutoff}} = 1.5$ | | | |
| Amati | $z < 1.5$ (94 GRB) | $0.81 \pm 0.29$ | $-2.03 \pm 0.67$ | $0.95 \pm 0.18$ | 86.14 | 225.76 |
| | $z \geq 1.5$ (147 GRB) | $1.13 \pm 0.15$ | $-2.09 \pm 0.38$ | $0.62 \pm 0.01$ | 139.62 | |
| Yonetoku | $z < 1.5$ (94 GRB) | $1.76 \pm 0.28$ | $-4.81 \pm 0.67$ | $0.65 \pm 0.03$ | 74.42 | 161.34 |
| | $z \geq 1.5$ (147 GRB) | $2.05 \pm 0.14$ | $-4.97 \pm 0.39$ | $0.30 \pm 0.01$ | 86.92 | |
| | | | Cutoff $z_{\text{cutoff}} = 2$ | | | |
| Amati | $z < 2$ (122 GRB) | $1.11 \pm 0.25$ | $-2.58 \pm 0.61$ | $0.94 \pm 0.13$ | 112.58 | 227.88 |
| | $z \geq 2$ (119 GRB) | $1.04 \pm 0.15$ | $-1.87 \pm 0.40$ | $0.60 \pm 0.01$ | 115.30 | |
| Yonetoku | $z < 2$ (122 GRB) | $1.89 \pm 0.23$ | $-5.01 \pm 0.57$ | $0.61 \pm 0.02$ | 98.55 | 166.54 |
| | $z \geq 2$ (119 GRB) | $2.04 \pm 0.14$ | $-4.87 \pm 0.41$ | $0.24 \pm 0.01$ | 67.99 | |
| | | | Filtered dataset ($F_\gamma \geq 2.6$ erg/cm$^2$/s) | | | |
| | | | Cutoff $z_{\text{cutoff}} = 1.5$ | | | |
| Amati | $z < 1.5$ (45 GRB) | $0.74 \pm 0.34$ | $-1.54 \pm 0.81$ | $0.65 \pm 0.29$ | 48.37 | 86.72 |
| | $z \geq 1.5$ (42 GRB) | $1.39 \pm 0.31$ | $-2.48 \pm 0.84$ | $0.58 \pm 0.13$ | 38.35 | |
| Yonetoku | $z < 1.5$ (45 GRB) | $1.74 \pm 0.28$ | $-4.41 \pm 0.67$ | $0.44 \pm 0.05$ | 39.66 | 73.58 |
| | $z \geq 1.5$ (42 GRB) | $1.90 \pm 0.18$ | $-4.27 \pm 0.51$ | $0.23 \pm 0.02$ | 33.92 | |
| | | | Cutoff $z_{\text{cutoff}} = 2$ | | | |
| Amati | $z < 2$ (56 GRB) | $1.20 \pm 0.35$ | $-2.48 \pm 0.84$ | $0.74 \pm 0.27$ | 56.27 | 86.47 |
| | $z \geq 2$ (31 GRB) | $1.33 \pm 0.33$ | $-2.38 \pm 0.94$ | $0.55 \pm 0.21$ | 30.20 | |
| Yonetoku | $z < 2$ (56 GRB) | $1.82 \pm 0.26$ | $-4.78 \pm 0.65$ | $0.45 \pm 0.05$ | 50.50 | 73.02 |
| | $z \geq 2$ (31 GRB) | $1.69 \pm 0.17$ | $-3.60 \pm 0.51$ | $0.16 \pm 0.01$ | 22.52 | |

In summary, high redshift bins always show better fitting than low redshift ones, regardless of the splitting method used, as noted from the $\sigma_{\text{int}}$ and AIC values. This indicates that both relations are more reliable for high redshift *Swift* GRBs than for low redshift ones. The large $\sigma_{\text{int}}$ and errors in the fitted parameters, A and B, of the Amati relation at low redshift suggest that the observational error does not encapsulate all the errors in the data. There is either an underestimation of the error or potentially new interesting physics at low redshift. The Yonetoku relation does not suffer from such problems. Additionally, the Yonetoku relation shows more consistent results for the two different datasets and is thus more reliable according to our results.

Since our results indicate that the Amati and Yonetoku relations are less reliable for the low redshift *Swift* LGRBs compared to the high redshift bursts, one might, for better accuracy and more targeted applications, exclusively use high redshift LGRBs to get better fits. The aim of the Amati and Yonetoku relations is to investigate the early universe, hence focusing on high redshift GRBs can cater more to this application of the two relations. We also note that in this case, the condition $F_\gamma \geq 2.6$ photons/cm$^2$/s, is not sufficient by itself, as many high redshift GRBs were eliminated, which affects the high redshift fittings. For better high-redshift investigations, one should account for the truncation of the dataset by methods such as the Efron-Petrosian method [46].

The question remains as to why we get relatively poorer fits from low redshift GRBs. One reason could be due to different origins of observed low redshift GRBs. As previously mentioned, it has been reported that there is an observed low redshift excess of LGRBs when one compares the redshift distribution of GRBs to the star-formation rate [41] [47]-[50]. Recent studies suggest that these LGRBs may emanate from other sources, like compact object mergers [48] [51]-[53], in which case the difference that we report regarding the applicability of the Amati and Yonetoku relations for low and high redshift might not be surprising.

## 4. Conclusion

In this work, the redshift evolution of the Amati and Yonetoku relations was investigated using *Swift* GRBs. A dataset of 241 LGRBs was compiled to carry out the investigation. In addition, a smaller subset of 87 LGRBs was formed that satisfied the condition $F_\gamma \geq 2.6$ photons/cm$^2$/s. Both datasets showed better fitting for the Yonetoku relation compared to the Amati relation. Two approaches were used to assess the redshift evolution of the two relations, the first one being a binning method where the datasets were split into 3, 4 and 5 equal bins and fit accordingly, and the second one was a redshift cutoff based split. Both approaches showed consistent results, with no clear redshift evolution being observed. However, our results indicate that the low redshift bins show poorer fits compared to the high redshift bins, thus making the Amati and Yonetoku relations more reliable for high redshift bursts and hence suitable for cosmological investigations.

## Conflicts of Interest

The authors declare no conflicts of interest regarding the publication of this paper.